\documentclass{sig-alternate}
\usepackage{enumitem}
\usepackage{etoolbox}
\usepackage{color,soul}


\begin{document}

\clubpenalty=10000 
\widowpenalty = 10000

\title{Extremely Large Images: Considerations for Contemporary Approach.}

\def\sharedaffiliation{
\end{tabular}
\begin{tabular}{c}}

\numberofauthors{4}
\author{
\alignauthor 
Slava Kitaeff \\
       \email{slava.kitaeff@icrar.org}
\alignauthor 
Andreas Wicenec\\
       \email{andreas.wicenec@icrar.org}
\alignauthor 
Chen Wu\\
       \email{chen.wu@icrar.org}
\alignauthor        
\sharedaffiliation
        \affaddr{International Centre for Radio Astronomy Research}\\
        \affaddr{University of Western Australia}\\
         \affaddr{M468, 35 Stirling Hwy, Crawley, WA, Australia}
\and
\alignauthor 
\alignauthor 
David Taubman\\
       \email{d.taubman@unsw.edu.au}
\alignauthor      
\sharedaffiliation
      \affaddr{School of Electrical Engineering and Telecommunications}\\
       \affaddr{The University of New South Wales}\\
       \affaddr{UNSW, Sydney, NSW 2052, Australia}
}

\maketitle

\begin{abstract}
The new wide-field radio telescopes, such as: ASKAP, MWA, LOFAR, eVLA and SKA; will produce spectral-imaging data-cubes (SIDC) of unprecedented volumes in the order of hundreds of Petabytes. Servicing such data as images to the end-user may encounter challenges unforeseen during the development of IVOA SIAP. We discuss the requirements for extremely large SIDC, and in this light we analyse the applicability of approach taken in the ISO/IEC 15444 (JPEG2000) standards.
\end{abstract}

\section{Introduction}

SIDCs from new radio telescopes that are currently in various stages of construction or commissioning -- Australian Square Kilometre Array Pathfinder (ASKAP) \cite{Cornwell2011}, Murchison Widefield Array (MWA) \cite{TINGAY}, LOFAR, MeerKAT, eVLA, ALMA -- are expected to be in the range of tens of GBs to several TBs. The Square Kilometre Array (SKA) Design Reference Mission, SKA Phase 1 \cite{SPDO2011}, defines at least one survey, namely the ``Galaxy Evolution in the Nearby Universe: HI Observations", for which the SKA pipeline will produce hundreds of SIDCs, of 70-90 TB each. In one year SKA Phase 1 is expected to collect over 8 EB of data. The data volumes for the full SKA are expected to be by at least an order of magnitude larger.

Even taking into account projected advances in HDD/SSD and network technologies, such large SIDCs cannot be processed or stored on local user computers. Visual exploration of such large volumes of data require a new paradigm for generating and servicing the higher level data products to the end-user and science processing HPC applications. In this discussion paper we make an attempt to define the functionality required to enable working with extremely large radio astronomy images, using JPEG2000 standard as a suitable example.

Currently, most radio astronomy data is stored and distributed in one of the three formats, namely: FITS \cite{CFITSIO-IO2011}; CASA Image Tables \cite{CASA}; and HDF5 \cite{Anderson2011, Anderson2010}. FITS and HDF5 are, in general, single self-describing files containing the image data as well as metadata. CASA, on the other hand, uses a different approach representing any data as a hierarchical structure of directories and files. CASA data is usually distributed as an archived file created by using common archiving software, such as e.g. \textit{tar} \cite{tar}. These formats provide both, portability and access to image data. Normally, SIDCs would be retrieved from an archive and stored on a local computer for exploration, analysis or processing purposes. Alternatively, a specified part of an image-cube (cutout) would be produced as a file or CASA Image Tables, and presented to the user as a download. If a coterminous regions are required, several cutout files would be produced and downloaded.

The International Virtual Observatory Alliance (IVOA) has developed the Simple Image Access Protocol (SIAP) \cite{IOVA} that defines a uniform interface for retrieving image data from a variety of astronomical image repositories. By using SIAP the user can query compliant archives in a standardised manner and retrieve image files in one or more formats, depending on the archive capabilities (e.g. FITS, PNG or JPEG), pretty much as outlined above. The resulting files can then be stored on a local computer or a virtual network storage device that is provided through VOSpace, which is another IVOA standard.

The purpose of this paper is to discuss the requirements for the use case of extremely large spectral-imaging data-cubes that are going to be produced during the operation of such new telescopes as SKA. We will discuss the limitations of IVOA SIAP for such images, and present the analysis of applicability of the approach taken in developing ISO/IEC 15444 standards to addressing similar requirements; and specifically JPIP standard that addresses interactive access to large images and their metadata.

\section{Use case for extremely large images}
The first thing to observe is that the extremely large SIDCs that are expected to be produced cannot be downloaded to a local computer. To explore even a fraction of an image, the cutout technique may generate very large files. Generation of such cutouts could present excessive demands on RAM, while the response time will be mostly limited by the server's I/O. The cutout service would require a significant additional capacity and performance from the storage that is already expected to be at the limit of what the technology will be able to offer. We suggest that a better approach would be to stream the required content directly from a suitably accessible compressed representation.

In many cases, images are not required at their full resolution or fidelity. It should be possible to access images at any of a multitude of reduced resolutions and/or reduced fidelities, according to need, all from a single master image. So called ``pyramid" organisations possess the desired multi-resolution accessibility attribute, but they involve about 30\% increase in storage requirements.

We also note that not all of the image might be required at the same level of quality. Particular regions of interest (ROI), e.g. containing objects of study such as a galaxy or a nebular, may need to be of much higher quality or resolution than others. Producing a cutout or many cutouts is a limited solution, as cutouts completely remove the surrounding area.  This is problematic because the surrounding area provides context for reconstructing the relationship between multiple objects in the field of view; moreover, the imagery within the surrounding area may be of interest in its own right. A much better approach in this case would be to have an adaptively encoded image, in which the regions of interest are encoded with higher fidelity/resolution than the surrounding areas.

Even combining such advanced techniques as multiple resolution/fidelity and adaptive encoding/transferring of ROI, the images can still be very large and require time to be transferred to the client. In the case of visual exploration of data, it would make sense to immediately transfer only the data that is required for displaying. Other parts of an image could be requested and transferred on demand. The protocol should be intelligent enough for such a use case.

It would also be very useful to support a progressive transfer. That is, the user should be able to see the whole image of the selected region queried as soon as a first portion of the data is transferred, while each successive portion of the data that is transferred should serve to improve the quality of the displayed imagery.  By contrast, many ``pyramid" techniques possess only multi-resolution access, without progressive transfer, so that higher quality representations must completely replace lower quality ones, leading to substantial inefficiencies and much higher transfer bandwidths. The client-server framework should be intelligent enough not to transfer more data than is necessary for displaying or processing the content that is of interest.

While there's still exist a wide spread believe that radio astronomy images, and data in general, can not be effectively compressed due to the noise-like nature of the signal, such view is not supported by evidence. In fact, there are several data and image compression techniques have been developed and trailed at ICRAR that have demonstrated that radio astronomy imaging data can be effectively compressed, and the error due to the compression can be controlled. Compression significantly reduces the cost of storage, operations and network bandwidth. However, it should be possible to access image regions, resolutions and qualities directly from the compressed representation.  If the imagery must first be decompressed, and then re-compressed to address a users needs, this will place unreasonable computational and memory demands upon the server, leading to a large latency in service time and limited ability to serve a variety of users. Ideally, decompression should occur only at the point where an image is to be displayed or used. Some usage cases can expect large ratios in compression; examples include visual data exploration, draft mosaicing, etc.  Other use cases may be less tolerant to loss of fidelity in the data, e.g. source finding. It follows that multiple levels of compression should be available: 
\begin{itemize}
\item high fidelity, potentially even numerically lossless compression, in which the decompressed image is either an exact reproduction of the original uncompressed image, or differs therefrom by considerably less than the intrinsic uncertainties in the imaging process; and
\item lossy compression, where the decompressed image exhibits higher levels of distortion that are considered acceptable in exchange for corresponding reductions in communication bandwidth or storage requirements.
\end{itemize}

As suggested by the last point above, distortion metrics need to be defined and made available to a user, so that the impact of lossy compression can be controlled.  Such metrics involve: 
\begin{itemize}
\item statistical characterisation of how the decompressed image can be expected to differ from the original image; and
\item measures of the impact that different levels of distortion can be expected to have on some specific purposes of data exploration, e.g. source finding.
\end{itemize}
The second point is especially important, given that much of new science is done at a very low signal-to-noise ratio (SNR).

\section{Case study: ISO/IEC 15444 (JPEG2000)}
In developing a contemporary protocol for working with extremely large astronomical images is useful to study how other communities have approached this problem. Indeed, large images are not unique to astronomy, though new telescopes such as SKA will be at the very extreme end of the spectrum. Medical imaging, remote sensing, geographic information systems, virtual microscopy, high definition video and other applications have long histories of development in the imaging domain. The large size of images is not the only similarity. Multi-frequency, multi-component, volumetric data sets, and metadata are common attributes in a range of existing imaging fields. A number of advanced image/metadata formats and access layer protocols have been developed over the years. Many of these are proprietary, such as MrSID (reference required). In this section we will discuss the ISO/IEC 15444 standard, also known as JPEG2000, as a particularly relevant open standards based solution, from which the astronomy community may learn.

JPEG2000 is an image compression standard and coding system. It was created by the Joint Photographic Experts Group committee in 2000 and published as an international standard, ISO/IEC 15444 \cite{Taubman02}. The standard was developed to address weaknesses in existing image compression standards and provide new features, specifically addressing the issue of working with large images. Considerable effort has been made to ensure that the JPEG2000 codec can be implemented free of royalties. Today, there is a growing level of support for the JPEG2000 standard, through both proprietary and open source software libraries.  JPEG2000 has been successfully used in a number of astronomy applications already, including the HiRISE (high resolution Mars imaging) project \cite{Powell2010} and JHelioviewer (high resolution Sun images) \cite{JHelioviewer}.

The considerations mentioned above led to several key objectives for the new standard. The new standard was expected to allow efficient lossy and lossless compression within a single unified coding framework as well as to provide superior image quality, both objectively and subjectively, at high and low bit rates. It was expected to support additional features such as ROI coding, a more flexible file format, and at the same time to avoid excessive computational and memory complexity, and excessive need for bandwidth to view an image remotely.

The main advantage offered by JPEG2000 is the significant flexibility of its codestream. The codestream obtained after compression of an image with JPEG2000 is scalable in nature, meaning that it can be decoded in a number of ways.  For instance, by truncating the codestream at any point, a lower resolution or signal-to-noise ratio representation of the image can be attained; moreover, the truncated representation remains efficient, in terms of the tradeoff that it represents between fidelity and compressed size.  By ordering the codestream in various ways, applications can exploit this so-called ``scalability" attribute to achieve significant performance benefits \cite{Taubman02}.

The main features which make JPEG2000 an attractive alternative to other image formats  currently used in radio astronomy are:
\begin{itemize}[leftmargin=*]
\item Superior compression performance. 
\item Availability of multi-component transforms, including arbitrary inter-component wavelet transforms and arbitrary linear transforms (e.g., KLT, block-wise KLT, etc.), with both reversible and irreversible versions.
\item Multiple resolution representation.
\item Progressive transmission (or recovery) by fidelity or resolution, or both.
\item Lossless and lossy compression in a single compression architecture. Lossless compression is provided by the use of a reversible integer wavelet transform and progressive transmission of a lossless representation provides lossy to lossless refinement.
\item Random codestream access and processing, also identified as ROI: JPEG2000 codestreams offer several mechanisms to support spatial random access to regions of interest, at varying degrees of granularity. These allow different parts of the same picture to be stored and/or retrieved at different quality levels.
\item Error resilience -- JPEG2000 is robust to bit errors introduced by communication channels, due to the coding of data in relatively small independent blocks within the transform domain.
\item Flexible file format: The JPX file format, in particular, allow for rich descriptions of colour and other metadata, and also allows images to be composed from any number of independently compressed codestreams.
\item Extensive metadata support and handling.
\item Support for volumetric image cubes, either through the specific set of extensions in Part 10 (a.k.a. ``JP3D") or by using the extensive set of multi-component transforms provided with Part 2 of the standard.
\item Interactivity in networked applications, as developed in the JPEG2000 Part 9 JPIP protocol. This feature of JPEG2000 deserves special consideration due to its utilisation in our proposed framework.
\end{itemize}

\subsection{JPIP}
JPIP is a client/server communication protocol, defined in Part 9 of the JPEG2000 suite of standards, officially entitled ``Interactivity Tools, APIs and Protocols;" it enables a server to transmit only those portions of a JPEG2000 image that are applicable to the immediate client's needs. Using an HTTP-based query syntax, together with TCP or UDP based transport protocols, JPIP enables the client to selectively access content of interest from the image file, including metadata of interest. This capability results in a vast improvement in bandwidth efficiency and speed when performing some very important and valuable image viewing tasks in a client/server environment, while reducing the storage and processing requirements of the client. The larger the images -- and the more constrained the bandwidth between client and server -- the greater are the benefits brought by JPIP.

JPIP clients access imagery on the basis of a so-called ``Window of Interest" (WOI).  The WOI consists of a spatial region, at a given resolution, within one or more image components in one or more underlying compressed codestreams, optionally limited to a desired quality level or amount of communicated data.  In advanced applications, the WOI may also be expressed relative to one or more higher level composited representations whose definition depends on metadata.  JPEG2000 enables the efficient identification and extraction of elements from the compressed codestream(s) that intersect with the WOI.   This means that from a single  compressed image, a user can remotely extract a particular region of the image, a large or small version of the image, a high or low quality version of the image, or any combination of these. JPIP can be used to progressively forward images of increasing quality, giving the client a view of the image as quickly as possible, which improves as rapidly as possible, along the direction of interest.

Such features are most desirable for extremely large radio astronomy images, which can hardly be used without examining the metadata and previewing the image at low resolution first, transferring only the selected parts of the image to a user's computer. This would normally require generating low resolution images, thumbnails and metadata and linking them all together in a database. In a system equipped with JPEG2000 and JPIP, however, it is only necessary to store a single file per image; lower resolutions and thumbnails can be extracted directly out of this high-resolution JPEG2000 ``master" image and downloaded. This removes the need to store, manage, and link images of different resolutions in the database, which can be cumbersome.

In a typical application, when the user chooses to view a particular image, only the resolution layer required to view the entire image on the screen need be transferred at first. Quality layers are downloaded progressively to give the user an image as quickly as possible. When the user zooms into a particular region of interest in the image, only that portion of the image is transferred by the server, and only at the resolution that is of interest. Again, the image can be transferred progressively by quality layers. The user can continue to zoom into the image until the maximum quality/resolution is reached, and pan across the image; each time, transferred content is limited to the area of the image being viewed. An interactive user might then scan across different images of a series, maintaining the same region and resolution of interest. Again, only the relevant content is actually transferred. The result is a dramatic increase in speed of viewing, and significant increase in the quality and efficiency of the viewing experience.

\subsection{JPIP Stream Type}
The JPIP standard allows three different types of image data to be transmitted between the server and client: 1) full, self-contained compressed images (typically, but not necessarily, in the JPEG2000 format); 2) tile data; and 3) precinct data \cite{Taubman}.

\textit{Full JPEG2000 Images}. For this data type the server sends to the client complete JPEG2000 images, at the requested resolution. The resolution level is selected to fit in the display window. Because the JPEG2000 images are self-contained, they do not require any additional metadata or headers during transmission; the images are simply sent to the client and the client decodes them.

\textit{Tiles}. Tiles are rectangular spatial regions within an image that are independently encoded. It can be useful to encode a large image as multiple independent tiles, but even huge images can be encoded as a single tile. A tile-based JPIP service is useful where numerous small tiles have been used during the encoding process; this allows the server to send only the relevant tiles to the client, for decoding. Because tile data is not a self-contained image, additional JPIP messaging headers are attached to convey to the client the contents of the messages.  It is worth noting that the use of small tiles reduces compression efficiency and can have a large adverse effect upon the service of reduced resolution imagery, since the effective size of the tiles within reduced resolutions can become very small.

\textit{Precincts}. Precincts are fundamental spatial groupings within a JPEG2000 codestream.  Unlike tiles, which represent independently coded regions in space, precincts are defined in the wavelet transform domain.  The detail subbands at each resolution level are partitioned into independently coded blocks, which are assembled into precincts.  Each precinct represents a limited spatial region within the detail bands that are used to augment the displayed imagery from a given resolution to the next.  Since precincts are defined in the transform domain, their contributions to the reconstructed imagery are overlapping.  This means that a server which sends the precincts that are relevant to a particular WOI is also sending some content that belongs to surrounding regions, whose extent is resolution dependent.  Precincts are the providers of ROI functionality in JPEG2000. The content of a precinct can be sent progressively, so as to incrementally refine the quality of the associated imagery.  Additional JPIP messaging headers are attached to the precinct data to convey to the client their contents. This image type is often the most efficient, as it requires the smallest amount of data to be transmitted; moreover, it is equally efficient at all spatial resolutions. An interesting potential mechanism for exploiting precincts within ASKAP and SKA applications, would be to use source finding algorithms to automatically generate a catalogue of the most relevant precincts, as part of the telescope pipeline. This would enable the selective storage of precinct data based on relevance (from lossy up to potentially numerically lossless), as well as the selective delivery of those precincts to a JPIP client; ``empty" parts of an image can be sent at much lower quality or resolution, saving the bandwidth and increasing the speed of fetching and viewing the data.

\subsection{JPIP Operation and Features}
The client application generates and sends to the server a properly formatted JPIP WOI request, containing information about the specific region of the image that the user wishes to view, along with the desired resolution, image components of interest and optionally explicit quality constraints -- alternatively, the client may request everything and expect to receive a response with progressively increasing quality. The JPIP server parses the request, calls the JPEG2000 library to extract the relevant image data, and sends back to the client a formatted JPIP response. When the response data is received, the JPIP client extracts the codestream elements and inserts them into a sparse local cache, from which the imagery of interest is decompressed, rendered and/or further processed on demand.  Importantly, JPEG2000 codestreams have such a high degree of scalability that any image region of interest can be successfully decoded from almost any subset of the original content on the server, albeit at a potentially reduced quality.  This means that decompression and rendering/processing from a local JPIP cache is an asynchronous activity that depends only loosely on the arrival of suitable data from the server.  To the extent that such data becomes available, the quality of the rendered/processed result improves.

Tile and precinct ``databins" are the basic elements of a JPEG2000 image used by JPIP. JPEG2000 files can be disassembled into individual finer elements, called \textit{databins}, and then reassembled. Each databin is uniquely identified and has a unique place within a JPEG2000 file. Full or partial databins are transmitted from the server to the client in response to a JPIP request. The JPIP client can decode these databins and generate a partial image for display at any point while still receiving data from the server.

JPIP provides a structure and syntax for caching of databins at the client, and for communication of the contents of this cache between the client and the server. A client may wish to transmit a summary of the contents of its cache to the server with every request, or allow the server to maintain its own model of the client cache by maintaining a stateful session. In either case, a well behaved server should reduce the amount of data it is transmitting in response to a JPIP request by eliminating the databins that the client has already received in previous transmissions. In this way, JPIP provides a very efficient means for browsing large images in a standards-compliant fashion.

Both precinct and tile databins have the property that they may be incrementally communicated, so that the quality of the associated imagery improves progressively.  JPIP also provides for the partitioning of metadata into databins, which can also be communicated incrementally.  This allows large metadata repositories to be organised and delivered on demand, rather than as monolithic data sets. Moreover, metadata can be used to interpret imagery requests and the image WOI can also be used to implicitly identify the metadata that is of interest in response to a JPIP request.

While databins are being transferred between the server and the client, they usually get split up into smaller chunks, called \textit{messages}. The JPIP server decides the JPIP message size. This flexibility to transmit partial databins enables one to vary the progressive nature of the data being sent to the client. If entire databins are sent, first for the lower resolution levels in the codestream and then for the higher resolution levels, the imagery pertaining to the requested WOI will be received in a resolution-progressive fashion; if messages from different databins at the same resolution level are interlaced, the data will be received by the client in a quality-progressive order. This flexibility allows applications to control the user experience, depending on the application requirements \cite{Taubman}.

\section{Remark on image formats}
We recognise that IVOA is not concerned with how the data is organised nor with the potential system performance, but rather concerned with interoperability. However, a complete ignorance of the image format by the access layer may limit the model for data interrogation. For example, the IVOA SIAP protocol can only provide access to JPEG2000 images, in accordance with a cutout prescription, while JPIP, developed with JPX/JP2 images in mind, is significantly richer. On the contrary, it would cumbersome to replicate JPIP functionality for FITS or CASA Image Tables, due to their internal data organisation. In our view, the development of a protocol and image interrogation model for extremely large images needs to be advanced together with the development of contemporary image data formats.

\section{Conclusion}

New telescopes such as SKA will produce images of an extreme size, that is well beyond the capabilities of current IVOA SIAP to provide an adequate performance and level of convenience. We believe a more advanced approach needs to be developed to provide interoperability to such large imaging data.

The ISO/IEC 15444 standard provides a possible framework that can be used to address the challenges of working with extremely large images.

\bibliographystyle{abbrv}
\bibliography{astro04-kitaeff}

\balancecolumns
\end{document}